# High-brightness single photon source from a quantum dot in a directional-emission nanocavity

Mitsuru Toishi[1,2, a]*, Dirk Englund[3,b]*, Andrei Faraon[4], and Jelena Vučković[1,c]

[1]Ginzton laboratory, Stanford University, Stanford CA 94305
[2]Sony Corporation, Shinagawa-ku, Tokyo, Japan, 141-0001
[3]Department of Electrical Engineering, Stanford University, Stanford CA 94305
[4]Department of Applied Physics, Stanford University, Stanford CA 94305
*Corresponding author: a) mitsuru.toishi@jp.sony.com, b) englund@fas.harvard.edu, c) jela@stanford.edu;

*These authors contributed equally to this work.

**Abstract:** We analyze a single photon source consisting of an InAs quantum dot coupled to a directional-emission photonic crystal (PC) cavity implemented in GaAs. On resonance, the dot's lifetime is reduced by more than 10 times, to 45ps. Compared to the standard three-hole defect cavity, the perturbed PC cavity design improves the collection efficiency into an objective lens (NA=0.75) by factor 6, and improves the coupling efficiency of the collected light into a single mode fiber by factor 1.9. The emission frequency is determined by the cavity mode, which is antibunched to g(2)=0.05. The cavity design also enables efficient coupling to a higher-order cavity mode for local optical excitation of cavity-coupled quantum dots.

## 1. Introduction

High-efficiency, high-indistinguishability single photon sources are needed in several quantum cryptography [1] and quantum computation [2, 3] schemes. Sources realized in photonic crystal (PC) devices embedded with quantum dots (QDs) are promising because they enable a large Purcell-enhanced emission rate into the cavity mode, which improves efficiency and photon indistinguishability [4-8]. In addition, QD-embedded photonic crystal nanocavities are also promising in several other applications requiring large light/matter interaction, including nonlinear quantum gates [9, 10] and photonic crystal lasers [11-13]. However, a major disadvantage is that the wide radiation pattern of photonic crystal nanocavities leads to low off-chip coupling efficiency. A poor mode overlap with single-mode-fiber further reduces the efficiency when it is necessary to fiber-couple the emission. Focused-ion-beam milling has been investigated as a technique to shape the far-field radiation pattern of a PC nanocavity by adding three-dimensional structure [14]. Although successful in changing the radiation pattern, the technique resulted in a significant degradation in the cavity's quality factor. Recently, we proposed a general method to perturb a given photonic crystal design with small added cylinders or slightly changed hole sizes to improve the far field pattern and output coupling efficiency [15]. By tuning positions and sizes of the perturbations to fit a desired mode, we can control the radiation pattern to improve out-coupling efficiency into a lens with given numerical aperture (NA), and enhance the mode overlap with a single-mode fiber for improved coupling. In this paper, we demonstrate a high efficiency single photon source consisting of an InAs QD coupled to a photonic crystal cavity with directional emission and high mode-overlap with a single-mode fiber (SMF). Using a collection lens with a numerical aperture of 0.75, the perturbed cavity design improved the collection efficiency by a factor of 6. Moreover, the fraction of the collected beam that could be coupled into a single mode was improved to 0.25 from 0.13. The improvement is possible because the perturbed cavity emission has a better overlap with the mode pattern of the SMF. The large extraction efficiency results in a large QD signal over background emission from the cavity. We measure antibunching of the signal collected directly from the cavity-QD system without the need for high-resolution spectral filtering. When the QD is tuned within one linewidth of the cavity, the cavity mode shows strong antibunching to a multiphoton probability of 0.05 compared to an equally intense Poisson-distributed source. From lifetime measurements on the cavity-coupled quantum dot emission, we estimate a Purcell factor enhancement exceeding 10.

## 2. Analysis of perturbed cavity

As our base cavity design, we consider a linear 3-hole defect cavity structure with side-holes shifted by 0.15 a, where a is the PC lattice constant and the unperturbed hole radius is 0.3a (Fig.1) [17]. We additionally reduced the radii of the holes directly above and below the cavity to 0.25 a. For improved extraction efficiency, the PC structure contains a Distributed Bragg Reflector underneath, as described in Ref.[9]. Theoretically, this structure offers a high quality factor (Q ~ 114,000), but has a wide radiation cone; from Finite-Difference Time Domain (FDTD) calculations, we estimate that only 30% of the emitted light is captured into an objective with NA=0.75. This estimate is derived by comparing the total emission into the light cone $|k_{\parallel,LC}|=\omega/c$ to the emission into the reduced cone given by NA*$|k_{\parallel,LC}|$, where $|k_{\parallel,LC}|$ is the in-plane field vector computed just above the PC slab, c is the speed of light in vacuum, and ω is the cavity resonance frequency. The far-field pattern of the unperturbed cavity, shown in Fig. 1(c), is far from the nearly Gaussian $HE_{11}$ mode pattern of a single-mode fiber [18]. As detailed in Ref.[15], we introduce a set of small perturbations to the base pattern to obtain a more directional far-field pattern that is also closer to the $HE_{11}$ mode. In the present application, these perturbations are arranged to scatter constructively in the far field above the center of the photonic crystal. The new structure is shown in Fig.1(a) and contains perturbations in the form of cylinders that are concentric around the original hole walls. Layer 2 (L2) perturbations are located at the sides of the hole cavity end-holes where the field amplitude is high, as shown in Fig.1 (a). Layers L3 and L4 have hole diameters that are roughly inversely proportional to the field amplitude (parameters are given in Fig.1). By optimizing the size and position of the perturbed points, the far-field pattern is more directional and has a mode pattern better matched to the fiber mode, with the coupling efficiency improved from 32.5% to 86.7 %. Here, we estimate the coupling efficiency by calculating overlap between the fiber mode and the cavity far-field emission after passing through a lens with NA of 0.75. and Gaussian profile of NA of 0.75.This is shown in Fig.1(d).

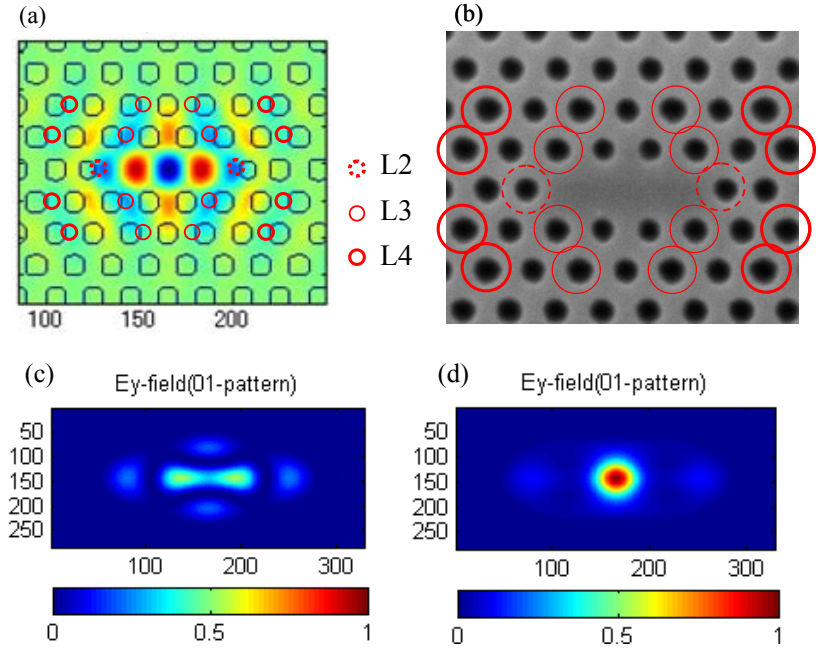

Fig. 1 (a) Field component (Ey) and positions of perturbation points, grouped in layers L2-L4, with diameters inversely proportional to the field amplitude |Ey|. (b) SEM pattern of perturbed PC cavity. The perturbation sizes at layers L2, L3 and L4 are 5nm, 10nm and 20nm, respectively. Calculated far-field pattern, at a distance above the sample surface, (c) without and (d) with perturbation points. By considering the mode overlap with the single mode fiber, these mode patterns predict an improvement in SMF coupling efficiency from 32.5 to 86.7%.

We fabricated these designs in a 165 nm-thick GaAs membrane containing a central layer of InAs QDs, using a combination of electron beam lithography and dry/wet etching steps [5]. The fabricated cavity is shown in Fig.1(b) and has a calculated mode volume of $V_{mode} = \int \varepsilon |\mathbf{E}|^2 / \max\{\varepsilon |\mathbf{E}|^2\} d^3\mathbf{r} = 0.8(\lambda/n)^3$, where $n = \sqrt{\varepsilon} = 3.5$ is the refractive index of GaAs at low temperature. We characterized the structure by photoluminescence in a confocal microscope setup[5]. The structures are maintained at a temperature of 10-30K in a helium continuous-flow cryostat. We measured a quality factor of 8500 from photoluminescence (PL) spectra, slightly below that of unperturbed three-hole defect cavities which had Q~11,000 (see Fig. 2 (b)).

To estimate the effect of the perturbations on the cavity out-coupling efficiency into a lens whose NA is 0.75, we measured the emission from the cavity when pumping above the saturation intensity of quantum dot lines (~10kW/cm$^2$) at a wavelength of 780 nm, above the bandgap of GaAs. The emission is collected by the objective lens and recorded on a spectrometer.

The Lorentzian cavity line-shape arises through the Purcell rate enhancement of the quantum dots. We will now describe how the cavity spectrum allows us to determine the coupling efficiency of the modified cavity compared to the base cavity. Suppose a quantum dot in the unpatterned bulk semiconductor emits at a spontaneous emission (SE) rate given by $\Gamma_0$. When coupled to a cavity mode, the SE rate of the saturated dot into the mode is modified to $\Gamma_0 \cdot F_{cav}$, where the Purcell factor into the cavity mode $F_{cav}(\vec{\mu},\vec{r},\lambda) = F_{c0}|\psi|^2 \cos^2\theta \cdot L(\lambda)$, where $\vec{r}$ and $\lambda$ are the position and wavelength of the QD; $L(\lambda) = (1+4Q^2(\lambda/\lambda_{cav}-1)^2)^{-1}$ is Lorentzian cavity spectrum; $F_{c0} = \frac{3}{4\pi^2}\frac{Q}{V'_{mode}}$ is the maximum Purcell factor; $V'_{mode} = V_{mode}/(\lambda/n)^3$ is the reduced mode volume; $|\psi| = |\vec{E}(\vec{r})/\vec{E}(\vec{r}_{max})|$ denotes the cavity field overlap at $\vec{r}$ normalized to the maximum field amplitude $|\vec{E}(\vec{r}_{max})|$; and $\cos\theta = \vec{E}(\vec{r})\cdot\vec{\mu}/|\vec{E}(\vec{r})||\vec{\mu}|$ gives the angle between the cavity field and QD dipole $\vec{\mu}$. The emission rate of the dot into all other modes is summed up as $\Gamma_{PC} = \Gamma_0 F_{PC}(\vec{\mu},\vec{r},\lambda)$. The total photon count rate collected after the lens from the saturated dot transition is then given by

$$\Gamma_{QD,lens}(\lambda,\vec{r},\vec{\mu}) = \Gamma_0(F_{cav}(\vec{\mu},\vec{r},\lambda)\eta_{cav} + F_{PC}(\vec{\mu},\vec{r},\lambda)\eta_{PC}) \qquad (1)$$

Here $\eta_{cav}$ and $\eta_{PC}$ are the coupling efficiency into the objective lens from the cavity and the averaged PC leaky modes. The total collected intensity is obtained by summing over all quantum dots inside the pumped and collected area $A$ on the sample,

$$\Gamma_{lens}(\lambda) = \sum \Gamma_{QD,lens}$$
$$\approx \int_A dA \int d\theta \, \Gamma_{QD,lens}(\lambda,\vec{r},\vec{\mu})\rho(\vec{r},\theta,\lambda) \qquad (2)$$

where $dA$ is a differential area. We estimated that the dots can be represented by a distribution function $\rho(\vec{r},\theta,\lambda)$ which is constant in $\vec{r}$ and $\theta$, and is given by the ensemble QD spectrum $\rho_{QD}(\lambda)$ in the wavelength $\lambda$. By combining (1) and (2), we obtain

$$\Gamma_{lens}(\lambda) \propto \rho_{QD}(\lambda)\Gamma_0 \left(\frac{1}{2}\int dA \, F_{c0}\eta_{cav}L(\lambda)|\psi|^2 + F_{PC}\eta_{PC}\int dA\right)$$
$$\approx \Gamma_0 \rho_{QD}(\lambda)\left(\frac{1}{2}A_{cav}F_{c0}\eta_{cav}L(\lambda) + F_{PC}\eta_{PC}A\right) \qquad (3)$$

where the cavity area $A_{cav} = \int dA |\psi|^2$ and the factor ½ in the first term comes from the fact that only half of the dots couple to linearly polarized cavity mode (for the second term, we have $\int d\theta \rho(\vec{r},\theta,\lambda) = \rho(\vec{r},\lambda)$). Thus we have

$$\Gamma_{lens}(\lambda) \propto \rho_{QD}(\lambda)(F_{c0}\eta_{cav}L(\lambda) + 2F_{PC}\eta_{PC}A/A_{cav})$$
$$= \rho_{QD}(\lambda)F_{c0}\eta_{cav}L(\lambda) + BG(\lambda), \qquad (4)$$

where $BG(\lambda) = 2\rho_{QD}(\lambda)F_{PC}\eta_{PC}(A/A_{cav})$ represents the background in the collected sample emission which is not related to the cavity mode. From $\Gamma_{lens}(\lambda)$ at the cavity resonance, where $L(\lambda) = 1$, we can then compare the coupling efficiencies of the perturbed and unperturbed structures:

$$\frac{\eta_{cav}(pert)}{\eta_{cav}(unpert)} = \frac{\Gamma_{pert,lens}(\lambda_{pert}) - BG_{pert}}{F_{c0,pert}\rho_{QD}(\lambda_{pert})} \frac{F_{c0,unpert}\rho_{QD}(\lambda_{unpert})}{\Gamma_{unpert,lens}(\lambda_{unpert}) - BG_{unpert}} \quad (5)$$

$$\approx \frac{\Gamma_{pert,lens}(\lambda_{pert}) - BG_{pert}}{\Gamma_{unpert,lens}(\lambda_{unpert}) - BG_{unpert}} \frac{Q_{unpert}}{Q_{pert}} \frac{\rho_{QD}(\lambda_{unpert})}{\rho_{QD}(\lambda_{pert})}$$

where the last step follows by assuming equal $V_{mode}$ for both unperturbed and perturbed structures, an assumption that is justified by our measurements of $\lambda_{cav}$ and FDTD simulations of $V_{mode}$.

In Fig. 2 (a), we plot the calculated outcoupling efficiency enhancement $\eta_{cav}(pert)/\eta_{cav}(unpert)$ for increasing layers of perturbations. We find that the collection efficiency reaches a maximum enhancement of factor 6 over the unperturbed L3 structure. This increase in the coupling efficiency is due to the increased directionality of the cavity emission, but probably also due to a lower fraction of emission being lost to material absorption. Since we do not know the exact material absorption coefficient (or the material-limited Q value), we are not able to distinguish experimentally between the two effects. This maximum efficiency is reached when adding the perturbations in layers L2-4. For this structure, the peak PL intensity is six times larger than that of the unperturbed structure. We also measured the fraction $\eta_{SMF}$ of the collected light that can be coupled into a single mode fiber (Thorlabs 980 nm single mode patch cable). As shown in Fig. 2 (a), we found nearly a factor two increase from 0.13 to 0.25. The simultaneous increases in out-coupling efficiency into the objective lens and coupling efficiency into the SM fiber represent substantial efficiency improvements for a range PC cavity devices.

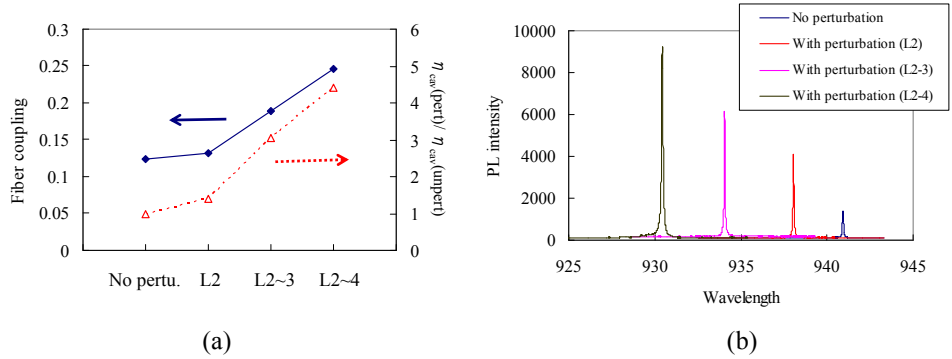

Fig. 2. (a) $\eta_{cav}(pert)/\eta_{cav}(unpert)$ (dashed red line) and fiber coupling rate (solid blue line) for different sets of perturbations. This PL intensity is measured with a spectrometer (coupled in free space). The fiber coupling rates are measured by comparing the PL intensity before and after a coupling to a single-mode fiber. (b) PL spectrum with and without perturbations.

The efficiency gain is interesting in the context of a single photon source, which we will now consider. We use a perturbed cavity that contains a single highly coupled quantum dot. The structure is excited with a Ti-Sapph laser producing 3.5ps pulses repeated at 80MHz. Its wavelength is tuned to a higher order-mode of the L3 cavity at 893nm. This higher-order mode is indicated in the PL spectrum in Fig.3 (a), which is obtained under optical excitation at 780 nm (above the GaAs bandgap). This pumping technique was proposed by Ref.[21] and allows us to selectively excite only dots that are spatially inside the cavity. To investigate the coupling between the QD and cavity, we tune both by temperature, with the QD shifting three times faster than the cavity. Figure 3(b) shows the PL spectra. When the cavity and the QD single exciton peak match at 22.5K, the QD peak is strongly enhanced via the Purcell effect. In this case, the exciton decay time changes by nearly factor six, as shown in Figs. 3 (c) and

(d). Compared to QDs in the bulk semiconductor, which have lifetimes ~ 600 ns, the Purcell enhancement is greater than factor 10. The reduced decay time can improve the indistinguishability of consecutive photons [4, 22, 23], and the spontaneous emission coupling efficiency into the cavity mode [6]. Furthermore, the QD repetition rate can be increase by up to 10 times compared to a QD in the bulk semiconductor.

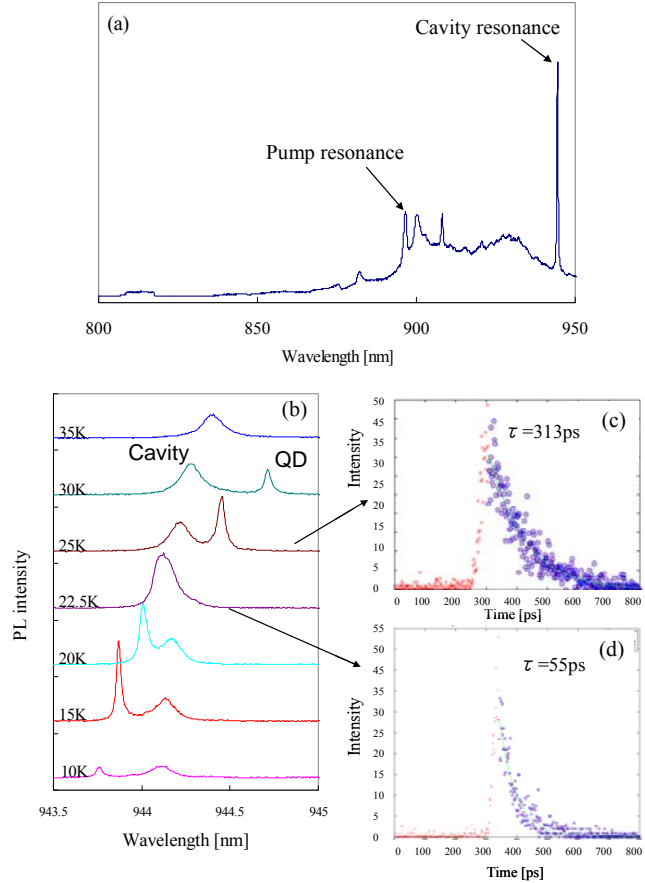

FIG.3 (a) Higher-order mode of PL spectrum .(b) PL spectra of PC nanocavities with at a range of temperatures. The QD emission is enhanced by coupling to the cavity. (c) and (d): Time decay spectra for temperatures 25.0K and 22.5K.

## 3. Single photon source demonstration

We evaluated the single photon source by approximating the autocorrelation function $g^{(2)}(t') = <I(t)I(t+t')>/<I(t)>^2$ using a Hanbury-Brown and Twiss (HBT) setup, where $t' = t_1 - t_2$ is the delay between events on the two detectors. First the QD emission is filtered to 0.2 nm with a grating setup. Figure 4 (a) shows the histogram of time correlation measurement. The antibunching of $g^{(2)}(0) = 0.04$ indicates that the multi-photon probability is suppressed to 4% below that of a classical source with Poisson-distributed photon statistics. We also measured the cross-correlation between the QD and cavity when the QD was detuned by -0.6nm from the cavity. The antibunching shown in Fig.3 (c) indicates that the cavity emission originates from the QD, as has been shown by other groups [24-26]. We speculate that the cavity is driven though a quantum dot dephasing process, as suggested recently by several groups [27, 28, 29]. We believe that the remaining $g^{(2)}(0) \sim 0.04$ results from background emission associate with cavity mode as well as repeated excitation of the QD within a single excitation pulse. Our detector resolution is 300ps and cannot resolve the temporal shape of the cross-correlation.

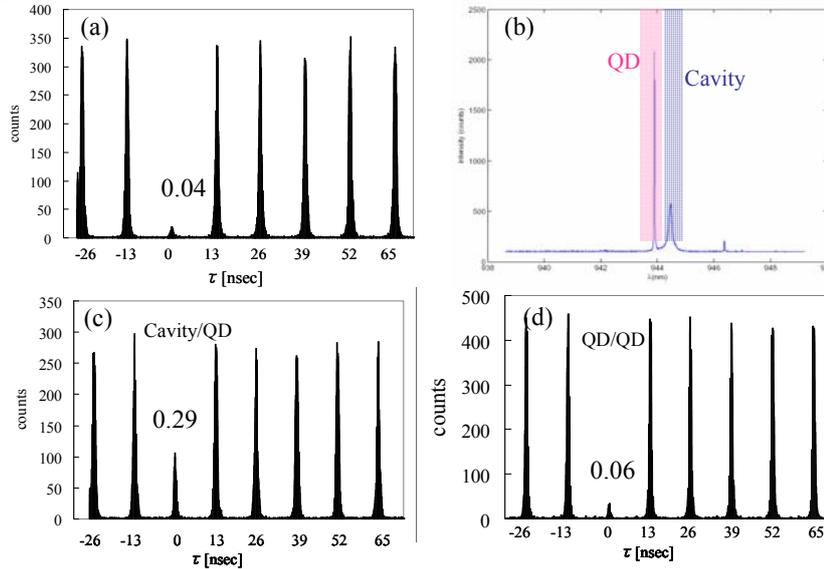

Figure 4 Auto- and cross-correlation measurements of the PC cavity emission. (a) Photon auto correlation histogram of emission using an HBT setup at 25K whose PL spectrum is shown Fig.3(b). The QD peak wavelength is selected by a grating. (b) The PL spectrum for cross correlation measurement. (c) Cross correlation measurement of QD and cavity. (d) Auto correlation measurement of cavity peak in (b).

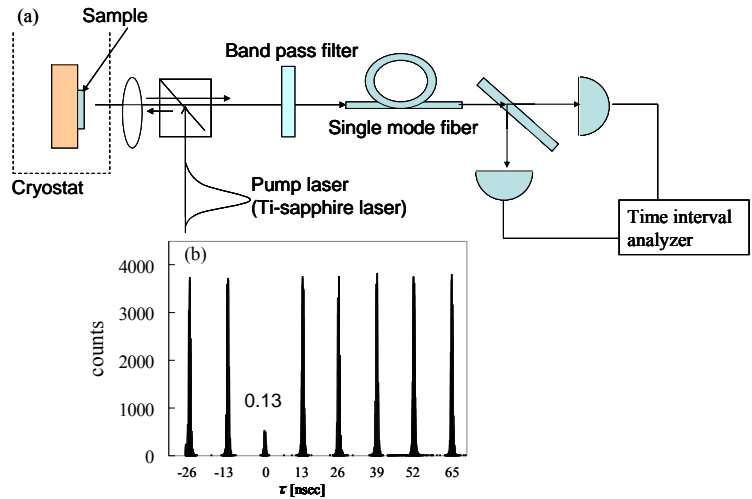

Figure 5 Fiber coupling experiment. (a) HBT setup with fiber coupling, using only a band pass filter to reject the pump beam. (b) Histogram of single photon with fiber coupling.

## 4. Conclusion

We have employed a new photonic crystal cavity design that greatly increases the directionality of the radiated field and its overlap to the mode pattern of a single mode fiber. We estimate that the coupling efficiency $\eta_{cav}$ into an objective with NA=0.75 is increased by factor 6, while the coupling efficiency into a single mode fiber is increased by up to factor 1.9 for the same design. We believe the increased coupling efficiency is due to higher directionality of the emission pattern and due to a lower fraction of the emission being lost to material absorption. A single quantum dot exciton coupled to the modified structure produces a train of single photons into the single mode fiber with far improved brightness, and without the need for high resolution spectral filtering. The cavity mode is strongly antibunched. The QD-cavity system thus represents a bright single photon source whose emission wavelength is determined by the cavity and rather insensitive to the potentially unstable QD emission wavelength. The directionality of the far-field radiation is interesting not only for out-coupling, but also for more efficient in-coupling of light.

**Acknowledgements**

Financial support was provided by the Army Research Office, ONR Young Investigator Award, Presidential Early Career Award, and the National Science Foundation.

We thank Nick Stoltz and Pierre Petroff of the University of California, Santa Barbara, for material growth.